 \documentclass[acmlarge,screen,nonacm]{acmart}

\usepackage{lipsum}
\usepackage{multirow}

\usepackage[utf8]{inputenc}
\usepackage{textgreek}

\usepackage{xcolor}
\usepackage{subcaption}

\newcommand{\rv}[1]{\textcolor{black}{#1}}      

\setcopyright{none}
\renewcommand\footnotetextcopyrightpermission[1]{}
\AtBeginDocument{%
  }

\begin{document}

\title{Modeling Trial-and-Error Navigation With a Sequential Decision Model of Information Scent}






\author{Xiaofu Jin}
 \affiliation{
   \institution{Aalto University}
   \city{Espoo}
   \country{Finland}
 }
\author{Yunpeng Bai}
 \affiliation{
   \institution{National University of Singapore}
   \city{Singapore}
   \country{Singapore}
 }
\author{Antti Oulasvirta}
\affiliation {
   \institution{Aalto University}
   \city{Espoo}
   \country{Finland}
 }

\renewcommand{\shortauthors}{Xiaofu et al.}

\begin{abstract}
Users often struggle to locate an item within an information architecture, particularly when links are ambiguous or deeply nested in hierarchies. \emph{Information scent} has been used to explain why users select incorrect links,
but this concept assumes that users see all available links before deciding. In practice, users frequently select a link too quickly, overlook relevant cues, and then rely on backtracking when errors occur.
We extend the concept of information scent by framing navigation as a sequential decision-making problem under memory constraints.
Specifically, we assume that users do not scan entire pages but instead inspect strategically, looking ``just enough'' to find the target given their time budget. To choose which item to inspect next, they consider both local (this page) and global (site) scent; however, both are constrained by memory. Trying to avoid wasting time, they occasionally choose the wrong links without inspecting everything on a page. 
Comparisons with empirical data show that our model replicates key navigation behaviors: premature selections, wrong turns, and recovery from backtracking.
We conclude that trial-and-error behavior is well explained by information scent when accounting for the sequential and bounded characteristics of the navigation problem.
\end{abstract}





\maketitle

\section{Introduction}
Navigating information architectures is challenging when links are ambiguous, overlapping, or buried in deep hierarchies.
The concept of \emph{information scent} has long explained why users follow incorrect links~\cite{pirolli1999information,fu2007snif}. However, most accounts assume that users inspect all available options in order to estimate scent and make decisions myopically at the current page. Analytical ``mathematical'' scent models such as CoLiDeS~\cite{Kitajima2000ColiDes}, CWW~\cite{blackmon2002cw}, and ACWW~\cite{blackmon2007cww} compute semantic relatedness and predict the top-rated link or expected clicks per page under full observability and deterministic choice. While precise, these models do not capture partial inspection, memory limits, or non-myopic planning. Heuristic simulation models such as SNIF-ACT~\cite{fu2007snif} and CogTool-Explorer~\cite{teo2011cogtool} introduced sequential evaluation of links and satisficing thresholds, but still rely on locally myopic strategies without a global probabilistic plan.
This gap leaves an opportunity to move beyond static or locally myopic formulations toward a \emph{sequential, resource-rational} account of navigation.

We extend the notion of information scent to capture how users make navigation decisions under uncertainty, including premature selections, backtracking, and revisiting previously explored options.
We approach this by framing scent-driven navigation as a sequential decision-making problem under memory limitations.
Our model assumes that users do not exhaustively scan entire pages. Instead, they adopt a strategy of inspecting ``just enough'' to identify a promising target within the constraints of their time budget.
In deciding what to inspect next, users weigh both local scent (on the current page) and global scent (across the site) under limited memory.
Because they aim to minimize wasted time, users sometimes commit to a link without fully inspecting all options, which leads to errors.

The model comprehensively reproduces trial-and-error behaviors established from empirical work:
\begin{itemize}
    \item \textbf{Partial scanning}: inspecting only a subset of items on a page before making a selection~\cite{byrne1999eye};
    \item \textbf{Backtracking}: returning to a previous page~\cite{scaria2014last};
    \item \textbf{Revisiting items}: going back to inspect earlier options~\cite{Jiang2014Searching};
    \item \textbf{Effects of information architecture}: capturing established empirical effects of task difficulties, hierarchy depth, and target position within the layout~\cite{habuchi2012characteristics,larson1998web, VANSCHAIK2001513, teo2011phd}.
\end{itemize}
%

We frame this navigation task a partially observable Markov decision process (POMDP), with explicit memory bounds, local and global scent, and a reward function that penalizes wasted time. And the simulated user behaviors emerge as resource-rational policy could be learned by reinforcement learning.
By comparing our model with empirical data, we demonstrate that it captures patterns of human exploration–exploitation in information architectures, including how users make mistakes and recover from them.

Taken together, our findings suggest that trial-and-error behavior can be effectively explained by information scent, once the sequential and bounded nature of the navigation task is taken into account.
This paper makes three contributions:  
\begin{enumerate}
    \item An extension of information scent theory to \emph{sequential decision-making under cognitive bounds}, explaining premature selections, backtracking, and revisits; 
    \item A computational model of navigation formulated as a \emph{POMDP with memory limits}, integrating both local and global scent into a non-myopic decision-theoretic framework grounded in resource rationality;
    \item An evaluation showing that the model reproduces hallmark human trial-and-error behaviors and effects of information architecture.
\end{enumerate}


\section{Related Work}

\subsection{Information Scent in Goal-Directed Navigation}

Information scent is the central construct in Information Foraging Theory (IFT). It is defined as “the user’s (imperfect) perception of the value, cost, or access path of information sources obtained from proximal cues”~\cite{pirolli1999information}. Such cues include link descriptors, images, contextual clues such as headings, or page arrangement~\cite{card2001infoscent}. Drawing on the \emph{marginal value theorem} (MVT)~\cite{Charnov1976MVT}, Pirolli and Card argued that users should leave a page once its perceived yield drops below that of alternatives, offering a normative account of how navigation balances local payoff against global opportunity~\cite{pirolli1999information}.

Empirical studies support this view: strong scent is linked to faster and more accurate navigation, whereas weak or ambiguous cues increase errors, backtracking, and abandonment~\cite{chi2001using,katz2003scent,scaria2014last}. At the same time, real navigation often departs from this ideal. Users rarely evaluate all options~\cite{byrne1999eye}, sometimes commit prematurely~\cite{scaria2014last}, and revisit earlier pages~\cite{Jiang2014Searching}. These behaviors suggest that scent guides decisions under bounded rationality, operating through partial and uncertain evidence rather than fully rational strategies.

\subsection{Computational Models of Scent-Driven Navigation}


Several computational models have sought to formalize or simulate how information scent guides navigation. 
Analytical models such as CoLiDeS~\cite{Kitajima2000ColiDes}, CWW~\cite{blackmon2002cw}, and ACWW~\cite{blackmon2007cww} operationalize semantic relatedness between user goals and link labels, predicting the most likely clicks or expected navigation costs. 
While mathematically precise, these models assume full observability and deterministic choice, and they are fundamentally \emph{myopic}: each step is evaluated in isolation, with the highest-scent link assumed to be chosen immediately, without considering the value of future options, the risk of error, or the cost of backtracking. As a result, they portray navigation as a multi-step comprehension process rather than a dynamic, experience-driven adjustment of belief (e.g., realizing “I am on the wrong path”). 

Heuristic simulation models provide a more process-oriented account. 
SNIF-ACT~\cite{fu2007snif} integrates scent into the ACT-R cognitive architecture to model sequential link evaluation and satisficing thresholds, while CogTool-Explorer~\cite{teo2011cogtool} builds on the same algorithm to offer a design tool for predicting user navigation. 
Although these models capture sequentiality, they inherit the same limitations. 
Their strategies remain locally myopic: users are modeled as following the link with the strongest immediate scent until satisficing is achieved, with no mechanism for probabilistic foresight, global planning, or strategic exploration. 
Backtracking, when it occurs, is treated as a consequence of failing the immediate choice, not as an anticipated or planned action.  

In summary, prior models have advanced understanding of scent-driven navigation but remain limited in their myopic assumptions, leaving open the opportunity for sequential, resource-bounded accounts that capture adaptive exploration under uncertainty.





\section{Modeling Scent-Based Navigation as a Sequential Decision-Making Problem}


The primary objective of this model is to reproduce human-like trial-and-error navigation in information architectures. Specifically, the model seeks to capture phenomena documented in empirical studies: how users adapt under varying task difficulties~\cite{habuchi2012characteristics,budiu2006navigation,blackmon2012scent}, how positional layout biases influence navigation choices~\cite{VANSCHAIK2001513,teo2011phd}, and how hierarchical structures shape efficiency and error rates~\cite{larson1998web,miller2004mesa}. We frame this problem as a partially observable Markov decision process (POMDP), reflecting that--like human users--the agent cannot directly observe the full structure of the environment or the true relevance of each option. Instead, it must form beliefs from limited cues, remember only a subset of them in memory, and plan actions sequentially under uncertainty. 

\begin{figure}[t]
    \centering
    \includegraphics[width=\textwidth]{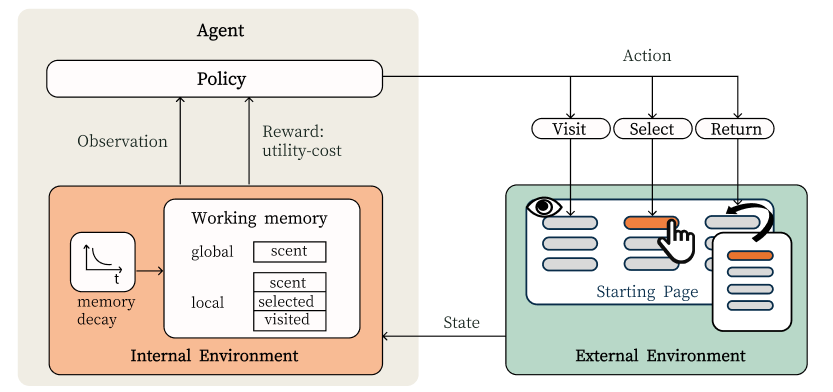}
    \caption{
    \rv{\emph{Model architecture.} The agent interacts with an external environment (a hierarchical information space, such as a menu) by choosing among three actions: visit, select, or return between pages to locate the target option. Each option has a latent information scent that becomes observable only upon inspection, and it is corrupted by perceptual noise. Observed cues are stored in working memory within the internal environment, where limited capacity and memory decay constrain retrieval. To handle uncertainty about where the target resides, the agent forms beliefs over states and selects actions according to its learned policy, receiving rewards (utility$-$cost) for reaching the goal and penalties for excessive steps. This resource-rational formulation frames human trial-and-error navigation as a sequential decision-making problem solvable with reinforcement learning.
    }}
    \label{fig:model}
\end{figure}

Figure~\ref{fig:model} illustrates the model architecture. The agent interacts with a hierarchical information environment by viewing a panel of local options while maintaining a small set of global memory traces. Each option has a latent information scent value~\cite{pirolli1999information}.
Scent becomes observable only when the option is inspected and is corrupted by Gaussian noise to reflect bounded perceptual precision. To simulate working memory limits, we restrict the number of cues that can be retained and apply decay so that unrehearsed traces fade over time~\cite{cowan2001magical,anderson1991reflections}. Given these bounded observations, the agent can choose to inspect another option, select a candidate, or return to a previous level. Rewards are given for successfully reaching the target, while small penalties are applied for each additional step. This cost-sensitive formulation follows the principle of resource rationality: humans optimize expected utility under constraints of limited cognitive resources such as memory and attention~\cite{simon1956rational,lieder2020resource}. Figure\ref{fig:wm} illustrates how working memory traces evolve over time in a sample navigation sequence.

\rv{
We now outline the structure of the remainder of this section. Section~3.1 examines four core \textit{computational principles} underlying the model: 
sequential decision-making under partial observability, a resource-rational objective, 
an information-scent based representation of options, and bounded memory capacity with decay.}
\rv{Section~3.2 then formalizes the POMDP, including the latent state representation, action space, transition dynamics, observation function, and reward. The reward function consists solely of a terminal success reward and a small per-step penalty. 
}

\subsection{Model Principles and Key Designs}

\subsubsection{Sequentiality}

Trial-and-error navigation is inherently sequential: users interleave information-gathering actions (e.g., visiting/inspecting links to reveal scent) with goal-directed actions (e.g., visit, select or return). Because each action changes what the user knows and where they can go next, the quality of a current choice depends on its long-run consequences. This motivates a sequential decision-making formulation.

Under policy $\pi$, the state-value and action-value functions are defined by the Bellman Q-value expectation equations:
\begin{equation}
Q^\pi(s,a) \;=\; \mathbb{E}\!\left[\, r_t \;+\; \gamma \, V^\pi(s_{t+1}) \,\big|\, s_t = s,\, a_t = a \right],
\label{eq:bellman-Q}
\end{equation}
where $s$ is the current state (e.g., the current page), $a$ is an action, $r_t$ is the immediate reward, $s_{t+1}$ is the next state, and $\gamma\in[0,1]$ discounts future rewards. 

Equations~\eqref{eq:bellman-Q} formalize two properties of navigation that static or purely myopic models cannot capture: (i) Value of information—visiting a link can be worthwhile even if it yields no immediate reward, because the revealed scent informs better choices later; and (ii) Path dependence—each action reshapes future opportunities and memory contents (via decay and capacity limits), so the best current action depends on anticipated downstream states.

\subsubsection{Resource Rationality}

We implement the principle of \emph{resource rationality} by introducing explicit limits on perception, memory, and effort. Following the resource-rational framework~\cite{simon1956rational,lieder2020resource}, we assume that humans maximize expected utility subject to cognitive constraints, rather than in an unconstrained optimal manner. In navigation tasks, this means that users seek to reach the target efficiently, but their decisions are bounded by noisy perception, limited memory, and the cognitive cost of additional actions.  

Formally, the agent chooses actions that maximize an expected utility function of the form:
\begin{equation}
    \pi^* = \arg\max_{\pi \in \Pi} \;
    \mathbb{E}[\,U(s,a)\,]
    \;-\;
    \mathbb{E}[\,C(s,a;\rho,t)\,],
\end{equation}
\rv{
where $U(s,a)$ is the task utility and $C(s,a;\rho,t)$ represents cognitive costs governed by resource-limit parameters $\rho$ (e.g., memory capacity, decay rate, perceptual noise) and elapsed time $t$.
This formulation states that the agent is not an unconstrained optimizer: each decision trades off utility against the costs imposed by human-like perceptual and memory limitations.
}

\rv{
In practice, these costs are instantiated in the POMDP via noisy scent observations, a capacity-limited global 
memory panel, and decaying memory traces with a retrieval threshold. Together, these mechanisms realize 
$C(s,a;\rho,t)$ within the observation and transition dynamics, so the learned policy is optimal only relative to human-like resource limits rather than an unconstrained ideal observer.
}
\rv{
Under these constraints, the optimal policy naturally exhibits human-like behaviors. Premature selections and backtracking emerge not as hand-coded heuristics but as boundedly optimal responses to noisy cues, forgetting, and action costs—unlike an ideal agent with perfect memory and no costs, which would exhaustively search without error.
}

\subsubsection{Information Scent}
Information scent refers to the local cues that users rely on--such as link labels, surrounding text, or semantic similarity--to estimate the value of a navigation option and decide whether to follow it~\cite{pirolli1999information}.
In the environment state, every option $i$ has a true scent value \rv{$\hat{\psi}_i$} defined as the semantic similarity between its label and the goal, following information scent theory~\cite{pirolli1999information}. We compute this similarity using a pretrained sentence-transformer \footnote{We use \texttt{paraphrase-multilingual-MiniLM-L12-v2}, a multilingual embedding model from the Sentence-Transformers library~\cite{reimers2019sentencebert,sentence-transformers}.}. 

Let $\mathbf{g}$ denote the embedding of the goal and $\mathbf{l}_i$ the embedding of option $i$’s label (each a 384-dimensional vector). 
The true scent is the cosine similarity lies in $[0,1]$. 
\rv{
\begin{equation}
    \hat{\psi}_i = \frac{\mathbf{g}\cdot \mathbf{l}_i}{\|\mathbf{g}\|\,\|\mathbf{l}_i\|}
\end{equation}}

However, the agent does not observe these values initially: all unseen options begin with a default scent of zero. 
Human judgments of semantic relevance are variable across trials, so we model observed scent as a noisy estimate of the true value:
\rv{
\begin{equation}
    \tilde \psi_i \sim \mathcal{N}(\hat \psi_i,\sigma^2)
\end{equation}}
\rv{This assumption is consistent with the general ACT-R principle that memory activations—and thus cue strengths—are noisy~\cite{anderson1998actr}. In our model, Eq.(4) captures this variability as uncertainty in information scent, aligned with resource-rational accounts of limited precision~\cite{lieder2020resource}.}
The specific value of $\sigma$ is not fixed by theory; instead we treat it as a tunable parameter within a plausible range (e.g.\ $0.01$--$0.1$ on the normalized $[0,1]$ scent scale), analogous to how diffusion parameters are calibrated in ACT-R~\cite{anderson1998actr}. 
A noisy scent \rv{$\tilde \psi_i$} is revealed only when the agent visits the option, reflecting the fact that users must inspect a label before assessing its relevance. 
Once revealed, scent values enter memory traces and are subject to decay, so that cues fade unless reinforced by repeated visits or clicks. 
This mechanism forces the agent to balance exploration (revealing new scents) and exploitation (following known strong cues). \rv{For clarity, Eq.~(3) and Eq.~(4) specifies how scent becomes observable when an item is inspected. These observed scent values enter the agent’s perceptual input but do not affect the reward function.}

\subsubsection{Memory Limit}
We model memory limitations through two mechanisms: capacity and decay.

\textbf{Working memory capacity.} Empirical evidence shows that working memory is severely constrained, with a practical limit of about 3--5 items~\cite{cowan2001magical,cowan2010magical}. 
We therefore restrict the active set of options to a few diagnostic candidates at any given time. This bottleneck enforces selective attention and naturally produces human-like trial-and-error when multiple links appear plausible.

\textbf{Memory decay over time.} 
Memory accessibility declines with time and interference, producing robust recency and frequency effects: recently and frequently encountered items are easier to retrieve~\cite{anderson1991reflections}. We capture this process as a resource-rational compression in which strong cues are retained while weaker ones fade~\cite{schooler2017adaptive}.

Each option $i$ is assigned a memory strength integrating three factors: its diagnosticity (information scent), repetition through views, and deeper processing through clicks:
\rv{
\begin{equation}
    M_i(t) = e^{-\lambda \Delta k_i(t)} \Big(b + a_s \hat{\psi}_i + a_v \sqrt{V_i(t)} + a_c \sqrt{C_i(t)}\Big)
\end{equation}}
Here, $\Delta k_i(t)$ is the number of steps since option $i$ was last seen, \rv{$b$ is a baseline activation term, $\hat{\psi}_i$} is the scent value, and $V_i(t)$ and $C_i(t)$ are cumulative views and clicks. Square-root scaling reflects diminishing returns from repetition~\cite{anderson1991reflections}, and clicks weigh more than views ($a_c>a_v$) to capture deeper processing~\cite{craik1972levels}.
The exponential term models time-based forgetting with half-life $H$, where $\lambda = \ln 2/H$. This provides a stable approximation of the power-law decay observed in human memory~\cite{anderson1991reflections,schooler2017adaptive}, though the formulation also admits a power-law variant consistent with cognitive architectures such as ACT-R~\cite{anderson1998actr}. \rv{In our model, each option maintains a memory-based cue representation whose strength is compared against the threshold~$\theta$::
\[
M_i(t) \ge \theta \;\Rightarrow\; \text{cue remains accessible}, 
\qquad 
M_i(t) < \theta \;\Rightarrow\; \text{cue is forgotten}.
\]
This comparison is performed independently for each stored cue: those with 
$M_i(t)\!\ge\!\theta$ remain accessible and contribute observable 
information, whereas cues below the threshold are treated as forgotten 
and no longer inform perception. This check is repeated at every step, so 
revisiting an item allows the agent to gain observable information again. In Section~3.2, 
we further reuse the activation values $M_i(t)$ to compute priority 
scores that determine which accessible cues enter the global-memory 
panel. Importantly, Eq.~(5) affects only the agent’s internal memory 
state—and thus the observations it receives—by gating cue accessibility 
and modulating priority scores; it does not alter the reward function.
}

\begin{figure}[t]
    \centering
    \includegraphics[width=1\textwidth]{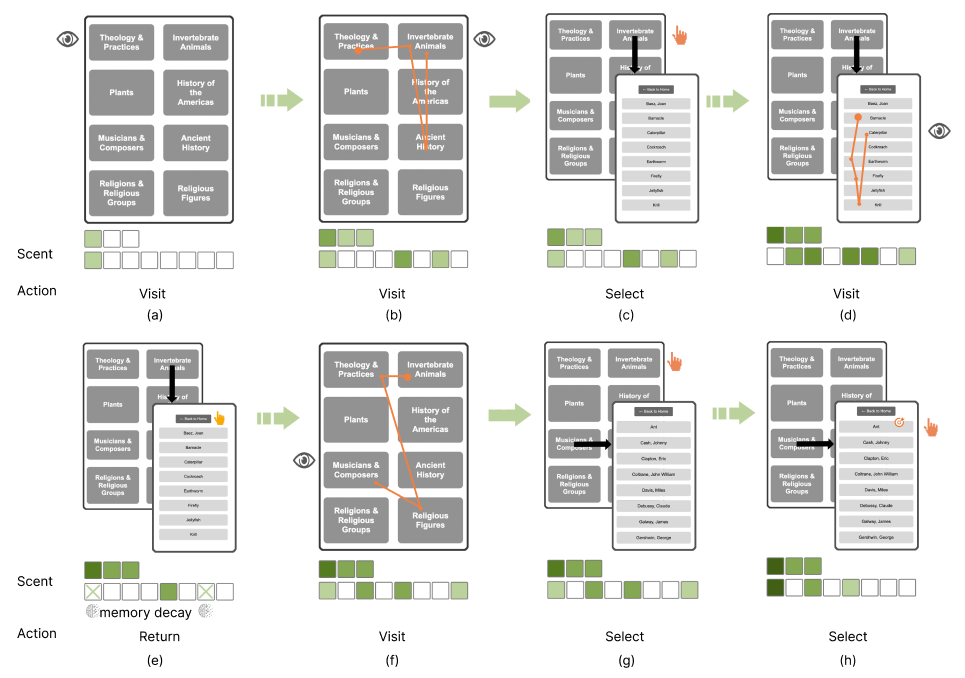}
    \caption{
    \emph{Illustration of the model’s decision process and state of working memory in a two-level menu task.} The agent navigates by choosing among \emph{visit}, \emph{select}, and \emph{return} actions, guided by perceived information scent (darker green = stronger cue). Each option has a latent scent; visits reveal noisy observations that update the agent’s belief. Scent is stored in working memory at two scopes: the \emph{top row} tracks the global top-\(k\) memory (\(k=3\) here) aggregated across visited pages; the \emph{bottom row} encodes the currently observed page.
    \textbf{(a–b)} On the top-level menu, visits expose option scents; the bottom row darkens for inspected items while the top row retains the strongest \(k\) cues across panels. \textbf{(c)} A strong local cue prompts \emph{select} (e.g., “Invertebrate Animals”), entering level two. \textbf{(d)} Further visits sample suboptions, sharpening local beliefs. \textbf{(e)} If local scents are weak, the agent may \emph{return}; and with memory decay, entries dropping below the threshold are forgotten ($\times$), so the agent forgets having visited the two squares. \textbf{(f–h)} The agent continues visiting and selecting the final target.
    }
    \label{fig:wm}
\end{figure}
\subsection{POMDP Formulation}
\label{sec:pomdp-formulation}

We formalize goal-directed navigation in information architecture as a POMDP 
$\mathcal{M}=\langle \mathcal{S},\mathcal{A},\mathcal{T},\mathcal{O},\mathcal{Z},R,\gamma\rangle$.
This provides a principled account of latent task state (where the target is and what has been explored), 
limited and noisy observations, and cost-sensitive decision making.

\textbf{State $\mathcal{S}$.} 
The full state contains all information about the hierarchical interface: the current layer, the index of the option in focus, the stack of previously selected parents, and the hidden target location. It also maintains traces of information scent and user history (past visits and clicks). As in real interaction, this full state is not directly observable to the agent.

\textbf{Action $\mathcal{A}$.} 
Actions are discrete navigation operations. A \textsc{Visit} action shifts focus to a candidate option, a \textsc{Select} action drills down into the selected option (or completes the task if it is the target), and a \textsc{Return} action moves back up one level. Together these actions capture the sequential structure of exploration, decision making, and error recovery in hierarchical interfaces.

\textbf{Observation $\mathcal{O}$.} 
At each step the agent perceives a bounded and noisy summary of the environment, consisting of a \emph{local panel} and a \emph{global-memory panel}. 
This design distinguishes between what is immediately visible on screen and what persists across steps in memory. \rv{Each option carries a scent cue: its observed value $\tilde{\psi}_i$ is computed via Eq.~(3)–(4).
All cues are then subject to the memory gate in Eq.~(5), which determines whether a trace is retained ($M_i \ge \theta$) or forgotten.}

\textit{Local panel.}
The local panel lists all options on the current layer. 
For each option $i \in \mathcal{L}_t$, the observation includes its currently revealed scent \rv{$\tilde \psi_i$ (or zero if never visited), along with discretized visit and click counts:
\[
x_i^{\mathrm{loc}} = \big(\tilde \psi_i,\, v_i,\, c_i\big).
\]}
If the current layer exceeds $N_{\max}$ items, we pad/clip to $N_{\max}$ rows for a fixed tensor interface (default $N_{\max}{=}12$). 
The local panel is fully observable in the sense that all items are listed, but the quality of their scent information depends on whether they have been inspected.

\textit{Global-memory panel.}
Independently of the current screen, we maintain a memory of the most diagnostic items seen so far. 
Each candidate $i$ is assigned a priority score:
\rv{
\[
q_i = \tilde \psi_i \cdot M_i, 
\qquad M_i = e^{-\lambda \Delta k_i}\Big(b + a_s \hat \psi_i + a_v \sqrt{V_i} + a_c \sqrt{C_i}\Big),
\] We denote the global candidate set at time~$t$ by $\mathcal{G}(t)$,
defined as the Top-$K$ items ranked by priority score~$q_i$\footnote{\rv{Visit and click counts are included only in the local panel because they reflect 
perceptual information available on the current screen and support within-screen behaviors such as detecting recent recurrence. The global-memory panel is purposefully more compressed: it summarizes longer-range diagnostic cues and does not store raw histories.}}.
For each $i \in \mathcal{G}(t)$, we expose a compact global feature 
$y_i^{\mathrm{glob}}$ maintaining its current scent estimate and 
normalized distance-to-goal:
\[
y_i^{\mathrm{glob}} = \big(\tilde \psi_i,\; d_i^{\mathrm{norm}}\big),
\]
where $d_i^{\mathrm{norm}}$ is a normalized distance-to-goal proxy, defined as
\[
d_i^{\mathrm{norm}} = \frac{\mathrm{cost}(s_t \!\to\! i)}{d_{\max}},
\]
with $\mathrm{cost}(s_t \!\to\! i)$ the shortest navigation path from current focus $s_t$ to option $i$ using 
\{\textsc{Return}, \textsc{Visit}, \textsc{Select}\}, and $d_{\max}$ an upper bound for normalization.}

\textbf{Transition $\mathcal{T}$.} 
State transitions are deterministic and mirror the structure of the information architecture. For example, a \textsc{Select} action pushes the current option onto the path stack and enters a child layer, while a \textsc{Return} action pops the stack and restores the previous layer. This design allows the model to simulate both progress toward the goal and backtracking when needed.

\textbf{Reward $R$.} 
The reward function balances success and efficiency. A large positive reward is given for reaching the target. Each step carries a small negative cost, which reflects the opportunity cost of time and effort and discourages fruitless detours. Formally, at step $t$ the agent receives
\[
R_t = 
\begin{cases}
+20, & \text{if the target is reached}, \\
-0.01, & \text{otherwise (per step)}.
\end{cases}
\] This cost-sensitive control implements the idea that people prefer efficient paths over exhaustive search. Backtracking is neither prohibited nor free: it becomes an effective strategy when local cues are weak, reproducing human-like error recovery. 

\begin{table}[h]
\centering
\caption{\rv{
\emph{Free parameters in the navigation model}. 
These parameters govern how the agent perceives, stores, and retrieves navigation 
cues. Their values were optimized using Bayesian optimization to improve 
model–human alignment.
}}
\label{tab:navigation-params}
\begin{tabular}{@{}llp{5cm}cccp{1.5cm}@{}}
\toprule
\textbf{Type} & \textbf{Name} & \textbf{Explanation} & \textbf{Range} & \textbf{Value} \\
\midrule
\multirow{1}{*}{Memory capacity} 
& $K_{glob}$ & Capacity of global memory panel (Top-K strongest cues retained) & $[3,5]$ & $4$   \\
\midrule
\multirow{1}{*}{Scent noise} 
& $\sigma$ & Standard deviation of Gaussian noise added to revealed scent values & $[0.01,0.1]$ & 0.08 \\
\midrule
\multirow{6}{*}{Memory decay} 
& $\lambda$ & Decay rate of memory trace; half-life $H{=}5$ steps & $(0,1]$ & $\ln 2/5$  \\
\cmidrule(lr){2-5}
& $b$ & Baseline memory strength (trace left after a single exposure) & $[0,1]$ & 0.50 \\
\cmidrule(lr){2-5}
& $a_s$ & Weight of scent strength in memory (diagnostic cues retained longer) & $[0,2]$ & 1.50 \\
\cmidrule(lr){2-5}
& $a_v$ & Weight of visit frequency in memory (recency/frequency effects) & $[0,1]$ & 0.80 \\
\cmidrule(lr){2-5}
& $a_c$ & Weight of clicks/selection (deep processing strengthens memory) & $[0,1]$ & 0.50 \\
\cmidrule(lr){2-5}
& $\theta$ & Retrieval threshold; below this memory is considered forgotten & $[0, \infty)$ & 1 \\
\bottomrule
\end{tabular}
\end{table}

\subsection{Parameterization and Training}
We approximate the solution of the POMDP using reinforcement learning. 
A policy network receives the bounded observation (Top-$K$ cues, local history) 
and outputs one of the atomic actions (Visit, Select, Return), with infeasible actions masked out. 
The policy is trained to maximize cumulative reward under sparse success signals and small step costs. 
Through repeated episodes, the agent learns to follow strong cues, to backtrack when evidence weakens, 
and to improve efficiency over time, thereby reproducing the practice effects observed in human navigation.



\begin{figure}[t]
    \centering
    \includegraphics[width=0.8\textwidth]{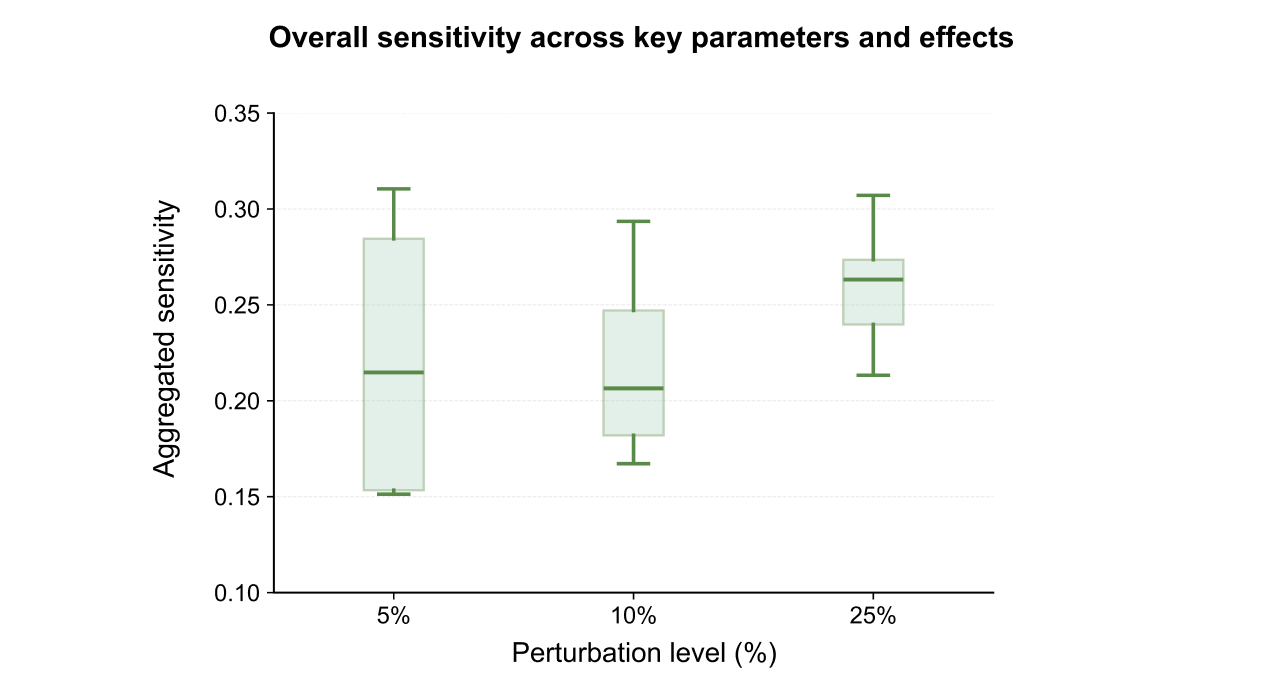}
    \caption{\rv{
    \emph{Overall parameter sensitivity across three perturbation levels (5\%, 10\, 25\%)}. Aggregated sensitivity values reflect the combined deviation in the three target effects when perturbing each parameter. The overall trend demonstrates that the model exhibits stable behavior under small perturbations and predictable, increases in sensitivity as perturbation levels grow.}}
    \label{fig:sensitivity}
\end{figure}

\rv{
We constrain all cognitive parameters to ranges informed by working-memory limits~\cite{cowan2001magical},
forgetting timescales~\cite{anderson1991reflections,schooler2017adaptive}, and perceptual uncertainty in
information scent~\cite{pirolli1999information} (Table~\ref{tab:navigation-params}). We then apply multi-objective Bayesian optimization over this space, targeting two human–model agreement metrics: a normalized delta-change metric for the difficulty effect and a relative-change score for the hierarchy effect. This yields an empirically calibrated parameter set that best matches human effects. Robustness is assessed by perturbing each optimized parameter by $\pm5\%$, $\pm10\%$, and $\pm25\%$ and recomputing the metrics; the model remains stable under small perturbations, with sensitivity increasing systematically as perturbation levels grow as Figure ~\ref{fig:sensitivity} shows\footnote{\rv{Aggregated sensitivity summarizes how strongly a parameter perturbation disrupts the three benchmark effects. For each parameter and perturbation level, we recompute: (i) the normalized delta-change metric for the task-difficulty effect, (ii) the relative-change score for the hierarchy-depth effect, and (iii) a qualitative position-effect consistency indicator (whether left/top remains better than right/bottom). Deviations from the baseline values on these three measures are rescaled to a common range and combined into a single sensitivity score; larger values indicate that small parameter changes distort multiple effects simultaneously.}}. Detailed parameter-wise effects are shown in Figure~\ref{fig:sensi-detailed}.
}

\section{Evaluation Method}

To evaluate whether our model reproduces established effects in information foraging, we designed proof-of-concept tasks using reconstructed HTML menu materials inspired by Blackmon et al. \cite{Blackmon2002, blackmon2007cww, teo2011phd}. Because the original stimuli are over two decades old and not fully reported, exact replication was not feasible, so we recreated the \emph{experimental conditions} as defined in prior studies and compared our results to the \emph{directional trends}, rather than absolute values\cite{habuchi2012characteristics, larson1998web, teo2011phd}.

We focused on three environmental factors known to influence navigation:

\begin{itemize}
    \item \textbf{Task difficulty.} Following Blackmon and Habuchi~\cite{Blackmon2002, habuchi2012characteristics}, we instantiated three levels: \emph{no-problem} (target link has a clear semantic advantage), \emph{competing} (distractors with similarly strong cues), and \emph{low-scent} (all links weakly related to the goal). Reported task items were used as anchors to align our instantiations with their operational definitions. For comparison, Habuchi et al.~\cite{habuchi2012characteristics} tested the same manipulation with 9 headings and 93 links derived from the Blackmon dataset, and found systematic increases in solution time and errors as difficulty increased.
    \item \textbf{Hierarchy depth}: We compared a two-level $8\times8$ menu (64 options) with a three-level $4\times4\times4$ menu (also 64 options). This parallels Larson and Czerwinski~\cite{larson1998web}, who showed that deeper hierarchies increase solution time and disorientation. While their menus contained more total items, holding the choice-set size constant allows us to isolate the effect of depth.
    \item \textbf{Target position}: To test spatial biases, we varied target placement (top vs.\ bottom; left vs.\ right) consistent with prior work. Visual search studies show that top/left targets are located faster~\cite{VANSCHAIK2001513}, and Teo~\cite{teo2011phd} demonstrated left–right asymmetries using an encyclopedia-style layout. To ensure comparability, we selected the same target items and categories reported in their study.
\end{itemize}

For each condition, we defined three task goals and evaluated the agent over 200 episodes per target. Performance was measured using established behavioral metrics: solution time (steps), click count, success rate, first-click accuracy, and lostness~\cite{smith1996lostness}. This design tests whether the model reproduces well-known effect directions: harder tasks require more steps, deeper hierarchies increase lostness, and top/left targets improve findability.

\rv{Finally, we conducted ablations on sequential foresight (via the discount factor~$\gamma$) and on the memory-decay and noise components, testing how each contributes to navigation efficiency and human-model alignment.}

\section{Results}



\rv{This section presents the behavioral patterns and empirical effects captured by our model. 
(1) We show how the agent reproduces key trial-and-error behaviors. 
(2) We evaluate its predictions against empirical effects of task difficulty, hierarchy depth, and target position. 
(3) We assess sequential foresight and the contributions of noise and memory components through ablation 
studies~$\gamma$.}

\subsection{Trial-and-error behaviors}

The model captures three core trial-and-error behaviors consistently observed in human navigation.

\paragraph{Partial scanning} The metric \emph{steps before first select} was consistently smaller than the number of items per page, showing that the agent inspects only a subset of available options before making a choice. This behavior mirrors human users’ tendency to stop scanning once a promising link is found~\cite{byrne1999eye}.

\paragraph{Backtracking} The \emph{return count} metric was greater than zero across conditions, reflecting systematic backtracking. A qualitative example is shown in Fig.\ref{fig:wm}d–e, where the agent retraces its steps after unsuccessful exploration, consistent with human navigation patterns\cite{scaria2014last}.

\paragraph{Revisiting items} The model exhibits two forms of revisiting. First, it returns to promising options once they are re-evaluated as better candidates (Fig.~\ref{fig:wm}b). Second, under memory decay it revisits previously inspected items whose traces have fallen below the retrieval threshold, in order to regain scent information (Fig.~\ref{fig:wm}e–f), consistent with human findings of revisiting in navigation~\cite{Jiang2014Searching}.


Beyond these trial-and-error behavior, we also reproduce the following effects of deign in information architecture reported in prior literature. 


\subsection{Effect of task difficulty}

\begin{figure}[H]
    \centering
    \includegraphics[width=\textwidth]{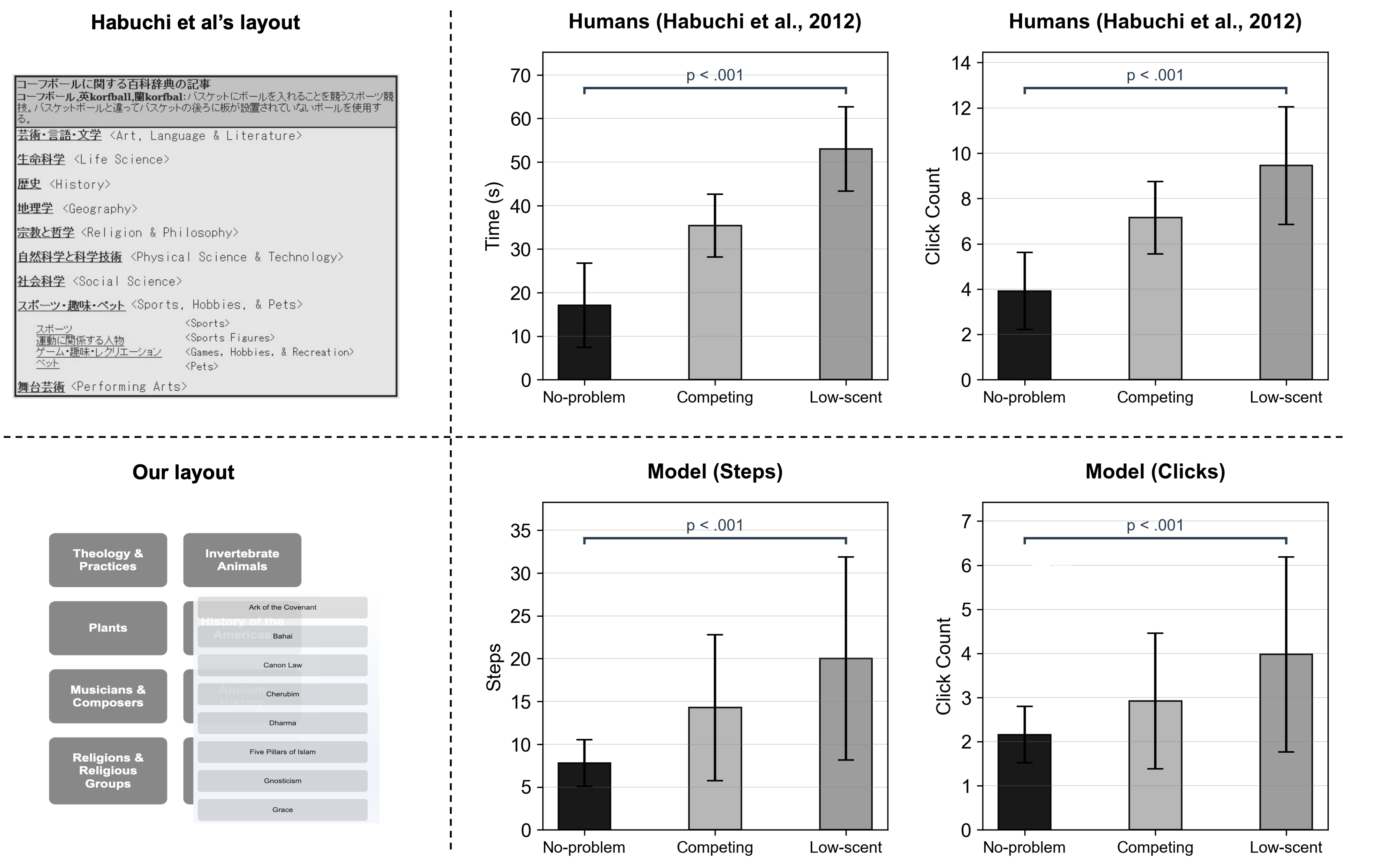}
    \caption{\rv{
    \emph{Navigation performance across task difficulties}. Human data (top) show that both solution time and click count increase systematically from no-problem to competing to low-scent conditions. Our model (bottom) reproduces this gradient, with more steps and clicks required under greater uncertainty. Beyond matching overall performance, the model also captures human-like behavioral dynamics: direct paths in no-problem tasks, trial-and-error loops in competing tasks, and revisits under low-scent conditions where weak cues and memory decay hinder efficient navigation.
    }}
    \label{fig:scent}
\end{figure}

We first examine task difficulty conditions defined by information scent. Habuchi et al. ~\cite{habuchi2012characteristics} reported significantly longer solution times and more clicks as difficulty increased
(all pairwise differences p < .001) as Figure~\ref{fig:scent} shows.

Beyond replicating the performance gradient (no-problem < competing < low-scent), our model also captured the dynamic behavioral patterns characteristic of human navigation. In the no-problem condition, agents typically followed direct paths. In the competing condition, trajectories often included trial-and-error loops, where plausible but incorrect links were explored before recovery. In the low-scent condition, agents not only exhibited trial-and-error loops but also revisits (Figure~\ref{fig:wm}). Because weak scent provided little guidance, agents required more steps, which increased the likelihood of memory decay and led to revisiting previously explored locations. These qualitative patterns are consistent with prior human observations~\cite{Blackmon2003Repairing,pirolli2007}.

\vspace{-20pt}
\subsection{Effect of hierarchy depth}

Larson and Czerwinski~\cite{larson1998web} demonstrated that deeper hierarchies substantially increase navigation difficulty. Comparing two-level (16$\times$32) and three-level (8$\times$8$\times$8) menu layouts, they found that three-level menus yielded longer solution times and higher lostness scores\footnote{\rv{The lostness score~\cite{smith1996lostness} is a standard metric of disorientation in hypertext navigation.}} than two-level menus ($p < .01$ and $p < .001$, respectively).



To test whether our model reproduces this effect, we compared a two-level 8$\times$8 layout with a three-level 4$\times$4$\times$4 layout, both containing 64 targets. The model replicated this depth effect: average steps increased from \rv{13.5} in the two-level hierarchy to \rv{25.6} in the three-level hierarchy ($p < .001$). Lostness scores likewise rose from \rv{0.19} to \rv{0.28} ($p < .001$).  

\begin{figure}[H]
    \centering
    \includegraphics[width=\textwidth]{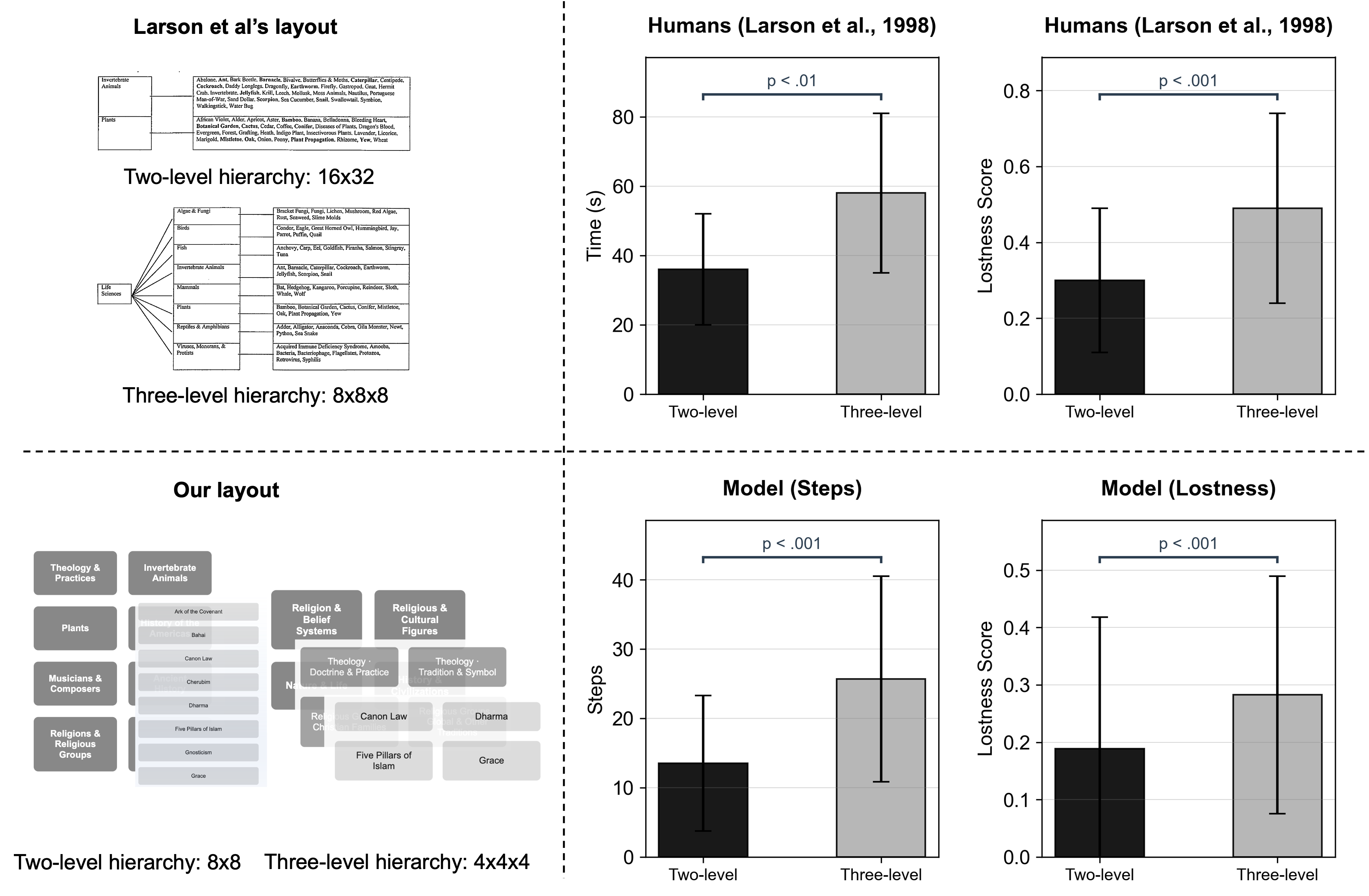}
    \caption{\rv{
    \emph{Effect of hierarchy depth on navigation performance}. Human data (top) shows that three-level hierarchies (8×8×8) significantly increased solution time and lostness compared to two-level hierarchies (16×32). Using our layouts with the same number of targets (bottom), the model reproduces this depth effect: three-level structures (4×4×4) required more steps and yielded higher lostness scores than two-level structures (8×8). This alignment indicates that deeper hierarchies amplify memory demands and trial-and-error exploration, making navigation less efficient for both humans and the model.
    }}
    \label{fig:hierarchy}
\end{figure}

\vspace{-200pt}

\subsection{Effect of target position}

\begin{figure}[H]
    \centering
    \includegraphics[width=0.85\textwidth]{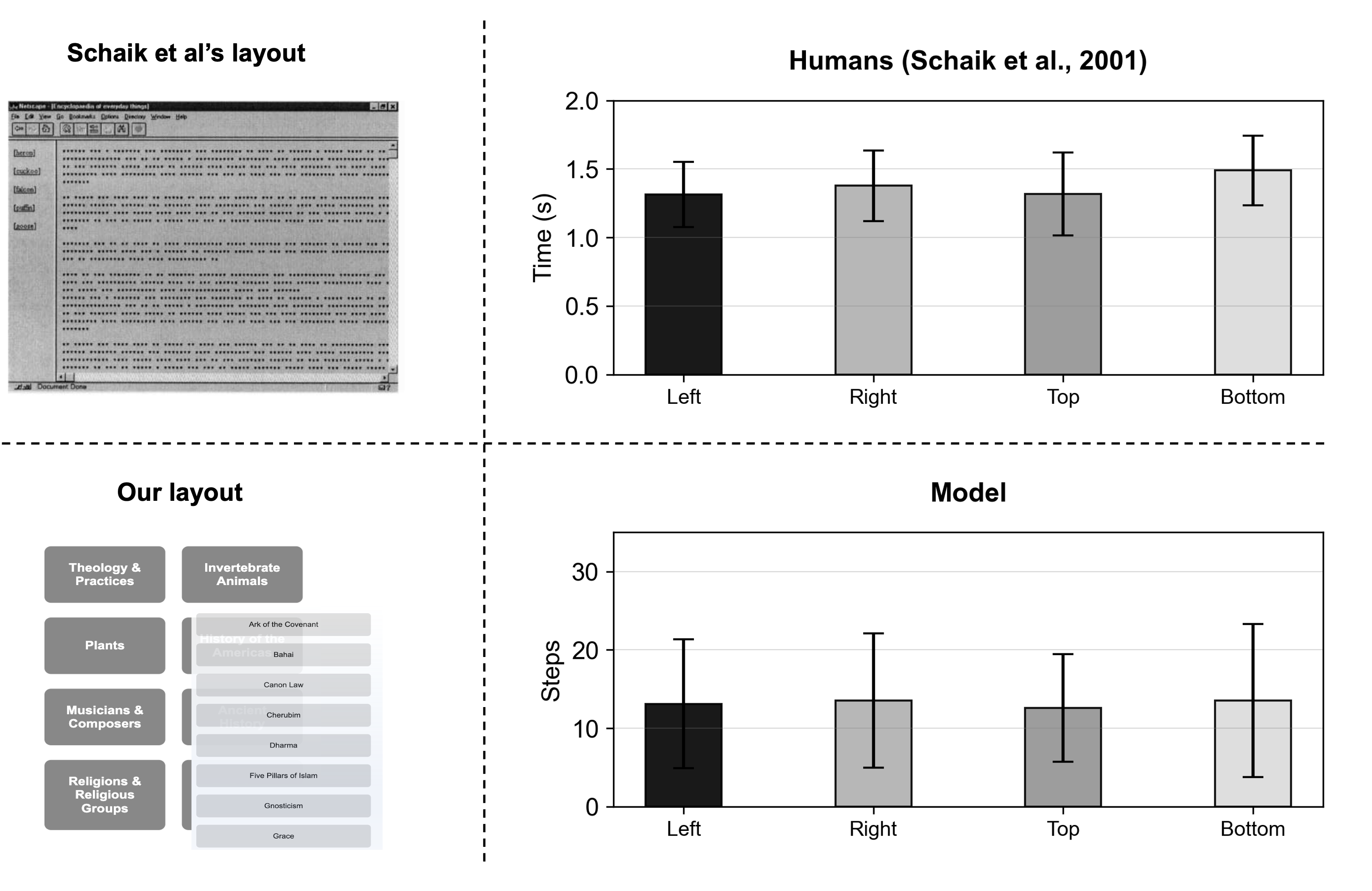}
    \caption{\rv{
    \emph{Effect of target position on navigation performance}. Human data (top) showed that targets placed on the left or top of a page were located more quickly than those on the right or bottom. Using our 8×8 menu layout (bottom), the model reproduces this positional bias: navigation required fewer steps when the target appeared at the left or top compared to the right or bottom. This alignment suggests that position effects in navigation can be explained as emergent consequences of bounded perception and sequential decision-making.
    }}
    \label{fig:position}
\end{figure}

Target position has also been shown to influence search performance. Schaik et al.~\cite{VANSCHAIK2001513} showed that links placed on the left or top of a page were located more quickly than those on the right or bottom (Fig.~\ref{fig:position}).


To test whether our model exhibits a similar position effect, we systematically varied the location of the target heading in our 8×8 menu. Consistent with human data, the model required fewer steps when the target appeared in the left or top positions compared to the right or bottom. Average steps were \rv{13.12} for left, \rv{13.52} for right, \rv{12.56} for top, and \rv{13.51} for bottom. These results replicate the human trend that search is facilitated when targets are positioned at the left or top of the interface.

Teo~\cite{teo2011phd} found that targets located on the left side were found more efficiently: participants required fewer clicks (left $\approx$\,1.12 vs.\ right $\approx$\,1.38) and achieved higher first-click success (left $\approx$\,90.4\%, right $\approx$\,66.4\%), with all differences significant ($p < .01$).

\begin{figure}[H]
    \centering
    \includegraphics[width=0.9\textwidth]{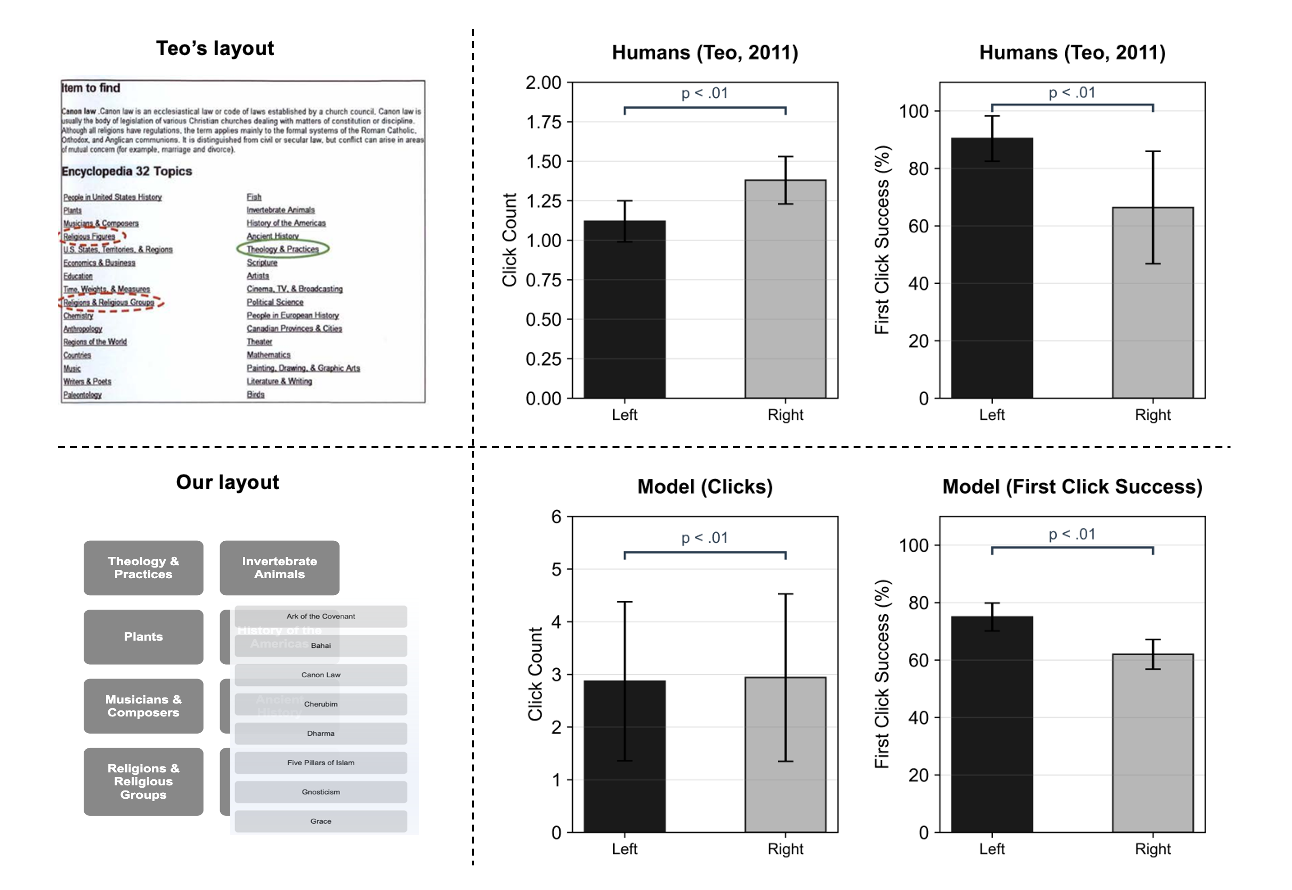}
    \caption{\rv{
    \emph{Effect of horizontal target position on navigation performance}. Human data (top) showed that targets located on the left side of a menu were found more efficiently than those on the right, with fewer clicks and higher first-click success rates. Using our 8×8 hierarchical layout (bottom), the model reproduced this positional effect: left-placed targets required fewer clicks and were more often selected correctly on the first attempt compared to right-placed targets. This demonstrates that spatial biases in human navigation can be captured as emergent outcomes of the model’s bounded search and decision-making process.
    }}
    \label{fig:position2}
\end{figure}

In our proof-of-concept setting using an 8$\times$8 hierarchical menu, we replicated this horizontal position effect. Although absolute values differed due to layout size, the model exhibited the same trend: left-placed targets were found more quickly and more often on the first click than right-placed ones, \rv{as Figure ~\ref{fig:position2} shows}. 

\rv{Beyond reproducing the empirical trend, the model’s positional preference can be understood as an adaptation to the statistical structure of the training environment. Following the view that human behavior is jointly shaped by cognitive limitations and the structure of the environment~\cite{Simon1969,Anderson1990,Oaksford2007}, we intentionally sampled more left/top configurations when generating training layouts. This choice reflects a recurrent structural regularity of real-world interfaces, which has emerged from designers adapting to established viewing patterns. Critical elements, such as brand identity and primary navigation targets, are overwhelmingly placed toward the upper-left region because this is the area where human visual attention is empirically highest, as shown by eye-tracking studies~\cite{Nielsen2008}. Under bounded perception and sequential evaluation, the learned policy therefore adapts to this environmental prior—producing a positional bias as an emergent consequence of ecological statistics in the task environment.}

\subsection{\rv{Ablation study}}

\begin{figure*}[t]
    \centering

    \begin{subfigure}[t]{0.48\textwidth}
        \centering
        \includegraphics[width=\linewidth]{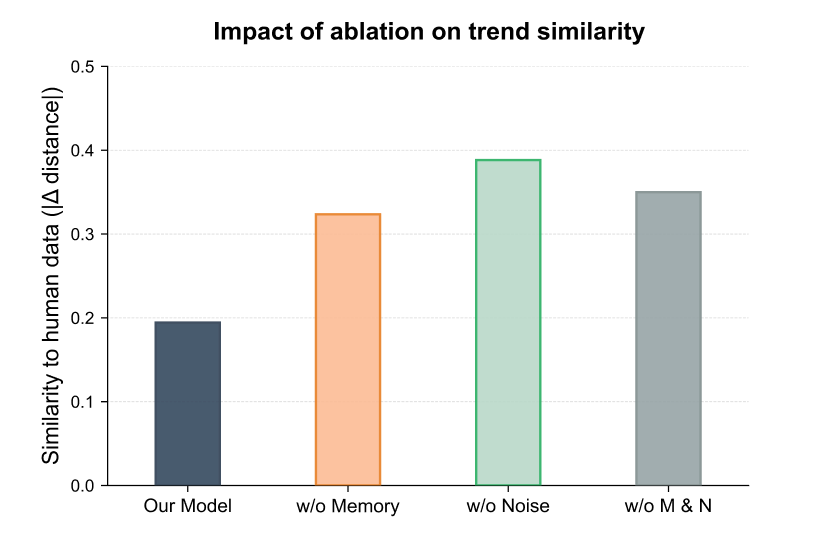}
        \caption{\rv{
    \emph{Ablation study of memory decay and noise components.} We evaluated the contribution of the memory decay and perceptual noise components by removing each module individually and jointly. The y-axis reports trend similarity to human data measured by absolute trend distance. Removing either component led to a substantial degradation in model–human alignment, with the strongest drop observed when removing perceptual noise.
    }}
        \label{fig:ablation-combined}
    \end{subfigure}
    \hfill
    \begin{subfigure}[t]{0.48\textwidth}
        \centering
        \includegraphics[width=\linewidth]{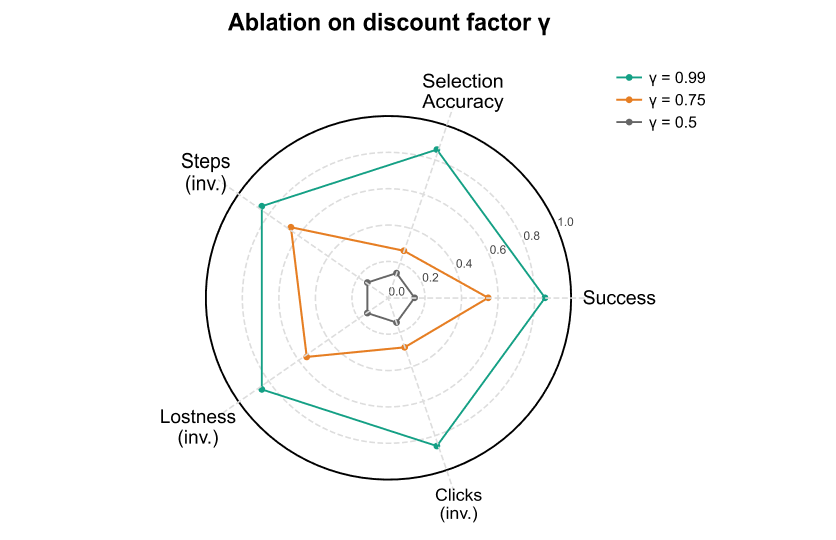}
        \caption{\rv{
    \emph{Radar plot illustrating the effect of the discount factor~$\gamma$ on five normalized performance metrics.} The plot compares the agent’s
behavior across $\gamma \in \{0.99, 0.75, 0.50\}$ along: Success, Selection Accuracy,
Steps~(inv.), Lostness~(inv.), and Clicks~(inv.). Metrics marked as “inv.”
indicate that lower raw values were inverted for visualization so that
axes share the same orientation for comparison.}}
        \label{fig:gamma-radar}
    \end{subfigure}

    \caption{Ablation Study Result Summary.}
    \label{fig:combined}
\end{figure*}


\subsubsection{\rv{Ablation: effect of memory and noise components}}

\rv{Figure~\ref{fig:ablation-combined} provides an overview of how removing components affects the model’s alignment with human behavioral
trends. Eliminating either the memory decay mechanism or the internal
noise term substantially increases the $\Delta$-distance from human data,
indicating that both components contribute to the qualitative
structure of navigation. Figure~\ref{fig:memory-ablation} shows how varying the memory decay threshold~$\theta$ affects the agent’s behavior. In our formulation, larger values of $\theta$ lead to more aggressive forgetting. As $\theta$
increases, the agent discards scent information associated with certain locations more quickly and must reconstruct its beliefs when needed, resulting in longer paths and more clicks. Return behavior and lostness also rise, indicating a shift toward more exploratory, trial-and-error
navigation.}
\rv{Figure~\ref{fig:noisestd} illustrates how the noise level~$\sigma$
affects the agent’s behavior. Increasing~$\sigma$ reduces the reliability
of scent discrimination, making the agent less certain about which option
is most informative. As a result, steps and clicks increase, and both return behavior and lostness rise, reflecting a
progressive shift toward less structured and more exploratory
navigation. These effects show that while some stochasticity is useful
for capturing human-like variability, excessive noise degrades the
agent’s ability to form stable preferences among competing actions.}





\subsubsection{Ablation: effect of discount factor $\gamma$}

The discount factor $\gamma$ (Bellman equation~\ref{eq:bellman-Q}) controls the planning horizon in reinforcement learning. High $\gamma$ (close to 1) values future rewards, encouraging cautious inspection of scent cues and reducing backtracking. Low $\gamma$ induces myopia: the agent favors immediate outcomes, selects prematurely, and neglects longer-term information. 

\rv{Figure~\ref{fig:gamma-radar} show that the
discount factor~$\gamma$ strongly influences the agent’s navigation strategy.
With a high $\gamma$ (0.99), the agent follows a stable long-horizon policy
characterized by high success, short paths, and low lostness. Reducing $\gamma$ shortens the
planning horizon and weakens this structure. At $\gamma=0.75$, navigation
remains functional but becomes less efficient, reflected in lower
accuracy and higher steps and clicks. At $\gamma=0.50$, the strategy
collapses into an impulsive mode characterized by premature commitments,
sparse inspection, and frequent mis-selections. Together, these results underscore that effective navigation requires \emph{sequential} foresight. This ablation validates our sequential formulation: goal-directed navigation cannot be explained by local heuristics alone but demands long-horizon planning under uncertainty.}

\section{Discussion}






Our model replicates the key trial-and-error navigation behaviors: partial scanning, backtracking, and revisiting. It also reproduces 3 effects of design reported in prior literature: (1) users take more time and clicks under competing or low-scent layouts, (2) deeper hierarchies increase navigation time and lostness, and (3) targets are located more easily when they appear at the left or top of a layout rather than the right or bottom.

Our model extends the notion of information scent by framing navigation as a sequential decision-making process under bounded resources with fewer handcrafted assumptions than possible earlier; thus, it can model trial and error behavior more broadly than possible before. 
A central contribution of this work is to demonstrate that navigation cannot be fully explained by static or myopic scent estimates. 
Sequentiality is key: users do not simply pick the link with the strongest immediate cue, but adaptively balance local and global information over time while managing uncertainty. 
By explicitly modeling this process, our account shows how errors and recoveries emerge naturally from bounded rational strategies, rather than being treated as failures of otherwise optimal choice. 
This sequential perspective connects information scent to broader theories of resource-rational cognition, where limitations in memory and attention shape adaptive exploration \cite{bai2026hierarchicalresourcerationalityexplains}.


\rv{Our findings have implications for both theory and design. For Theory, our model links Information Scent with decision-theoretic accounts of navigation, offering a 
resource-rational framework that formally incorporates cognitive constraints such as memory decay and uncertainty. This extends existing IFT models with a quantitative, predictive tool that generalizes to other sequential tasks (e.g., menu selection, search, information retrieval).
For Design Practice, the model provides a computational lens on where and why users ``get lost'' in complex information structures~\cite{pirolli1999information}. By quantifying uncertainty and expected information gain, it offers actionable guidance for improving navigation support, as detailed below.}

\begin{enumerate}
    \item \rv{\textbf{Computational evaluation of Information Architectures (IA):} The model serves as an efficient computational complement to traditional evaluation methods such as tree testing~\cite{Eduard2025Validation}by predicting backtracking, path inefficiencies, and confusion points~\cite{rosenfeld2015information,tullis2013measuring}, enabling early identification of structural issues without running full studies.}
    
    \item \rv{\textbf{Identifying scent and labeling failures:} By explicitly computing perceived utility for each category, the model reveals where scent collapses or labels induce high uncertainty~\cite{chi2000scent}, supporting precise revisions in wording and grouping.}
    
    \item \rv{\textbf{Predicting lostness~\cite{gwizdka2007implicit,UXMatters2025NavigationCost} and intervention timing:} By modeling how memory decay and ambiguity accumulate, the model predicts *when* and *where* disorientation emerges~\cite{gwizdka2007implicit}, guiding the timely introduction of previews~\cite{zhang2001implications}, visited-state indicators~\cite{Damien2022Supercharging}, and breadcrumbs~\cite{Nielsen2020Breadcrumbs}.}
    
    \item \rv{\textbf{Supporting optimal stopping decisions and branch abandonment:} The agent’s expected-gain calculations identify where users abandon too early or persist too long~\cite{pirolli2007,fu2007snif}, offering guidance on restructuring categories or reinforcing scent.}
    
    \item \rv{\textbf{Enabling proactive and personalized design adaptation:} Because the model tracks how scent ambiguity, policy entropy, and memory decay increase uncertainty, it can indicate *when* and *for whom* additional cues are needed. This supports adaptive designs—for example, stronger visual cues for users with lower working-memory capacity~\cite{Xiaofu2024Explore} or more prominent navigation aids when uncertainty rises~\cite{hao2025uncertainty}.}
\end{enumerate}


\rv{A methodological limitation of our evaluation is that we do not align the model with raw human navigation trajectories. In practice, trajectory-level comparison is not feasible for hierarchical IA tasks: individual paths vary widely and are strongly influenced by unobserved factors such as prior familiarity, motor habits, or search styles. For this reason, prior models of menu and web navigation have also evaluated against aggregate behavioral signatures (e.g., lostness, partial scanning) rather than attempting to predict individual click sequences. Moreover, no publicly available dataset provides process-level traces for goal-directed hierarchical navigation.
Given these constraints, we adopt the standard cognitive-modeling strategy of validating against 
well-established benchmark effects, which offer controlled, theoretically grounded testbeds that isolate the mechanisms under study. The modern-interface case studies we include (Fig.~\ref{fig:case}) serve as demonstrations of external applicability rather than quantitative validation: real-world websites differ in visual salience, semantics, and interaction affordances, and meaningful human–model comparison on these interfaces would require new matched datasets.
}

\rv{Future work could enrich this direction by collecting process-level logs and 
clustering human trajectories into typical strategy patterns, enabling 
qualitative and empirical comparison between human and model strategies and 
connecting our framework to recent computational models of attention and 
interaction behavior \cite{bai2024heads, shi2025chartist}.} 
\rv{In addition, our current model does not capture users’ prior knowledge, semantic associations, or individual differences in memory bounds. Incorporating such factors offers an important avenue for improving generalizability and modeling human variability.}

\section{Conclusion}
We modeled trial-and-error navigation as a sequential, resource-rational decision process. Our results show that behaviors often seen as errors--premature selections, backtracking, and revisits--emerge naturally once perception, memory, and effort limits are taken into account. Ablation studies highlight the importance of sequential foresight: only with long-horizon planning (high $\gamma$) does the agent behave efficiently, while reduced foresight leads to myopic, error-prone exploration. By extending information scent into a resource-rational framework, our model explains and predicts how task difficulty, hierarchy depth, and target position shape navigation performance. This approach offers a principled basis for understanding and improving information architectures without hand-crafted heuristics.

\begin{acks}
To Robert, for the bagels and explaining CMYK and color spaces.
\end{acks}


\bibliographystyle{ACM-Reference-Format}
\bibliography{sample-base}


\appendix
\clearpage

\begin{figure}[t]
    \centering
    \includegraphics[width=0.75\textwidth]{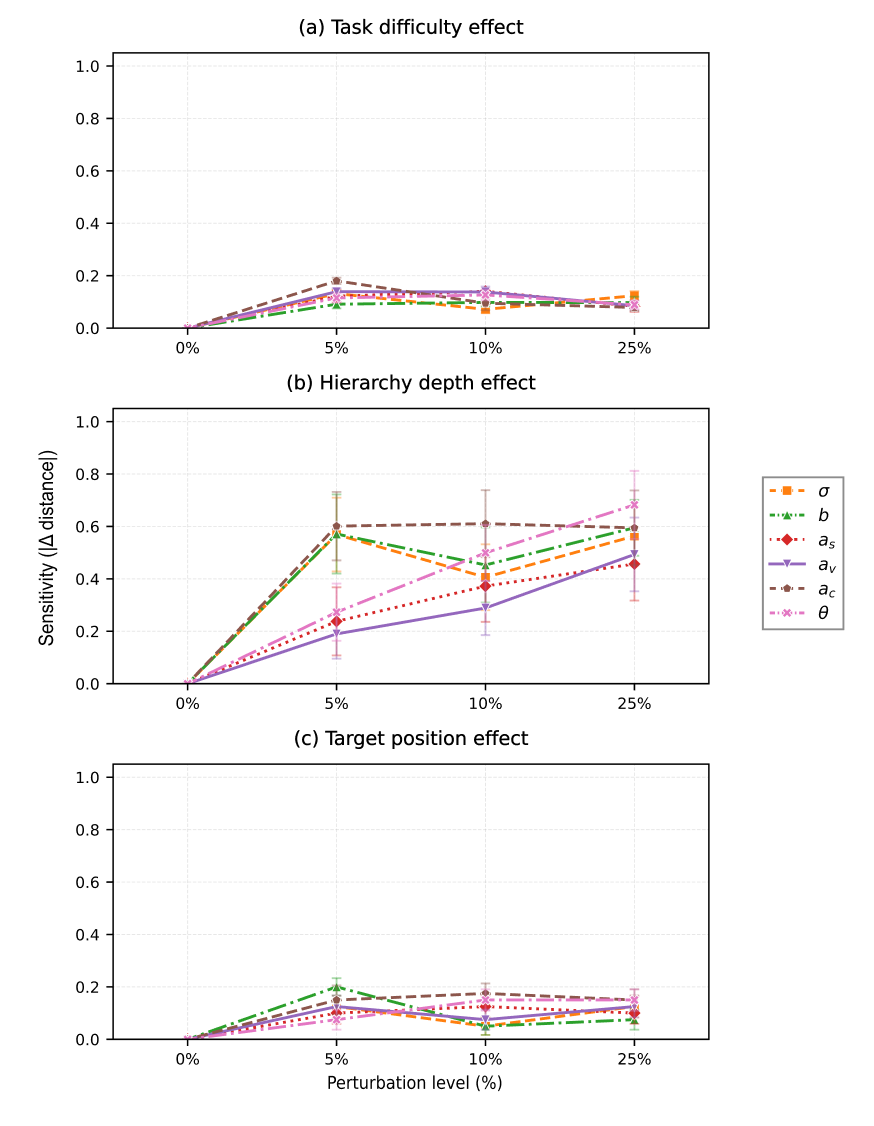}
    \caption{\rv{
    \emph{Parameter sensitivity analysis across three effects.} (a) Task difficulty effect, (b) hierarchy depth effect, and (c) target position effect. Each panel shows the change in trend similarity (Δ distance) under systematic parameter perturbations ( 5\%, 10\%, 25\%) relative to the optimized model. Different colored lines represent different model parameters. Error bars indicate $\pm$1 standard deviation across trials.}
    }
    \label{fig:sensi-detailed}
\end{figure}

\begin{figure}[t]
    \centering
    \includegraphics[width=1.0\textwidth]{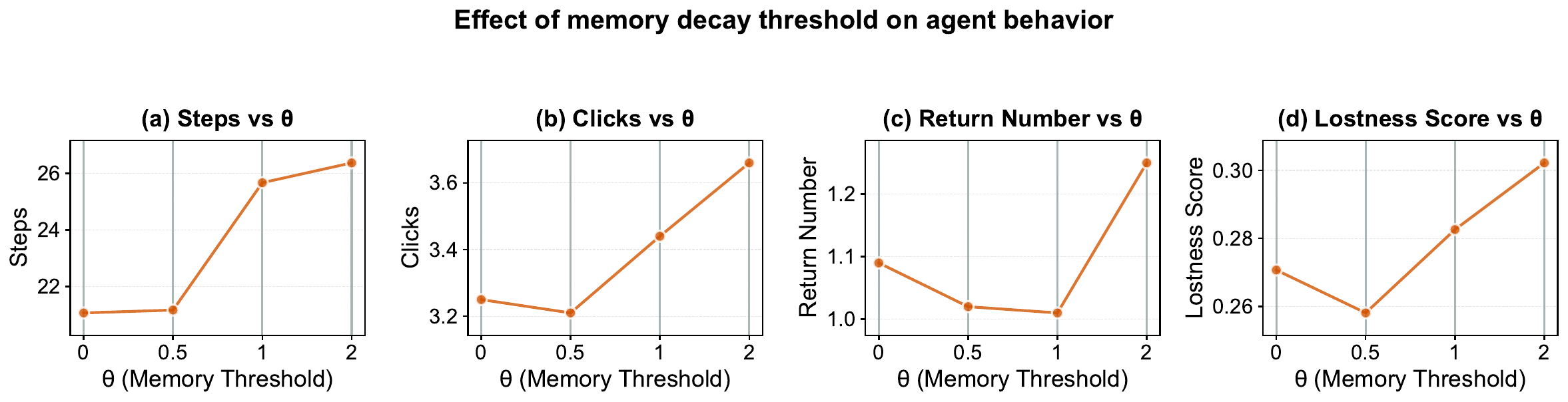}
    \caption{\rv{
    \emph{Effect of the memory decay threshold~$\theta$ on navigation
behavior.} Panels show how varying~$\theta$ influences:
(a) steps, (b) clicks, (c) return number, and (d) lostness score.}}
    \label{fig:memory-ablation}
\end{figure}

\begin{figure}[t]
    \centering
    \includegraphics[width=1.0\textwidth]{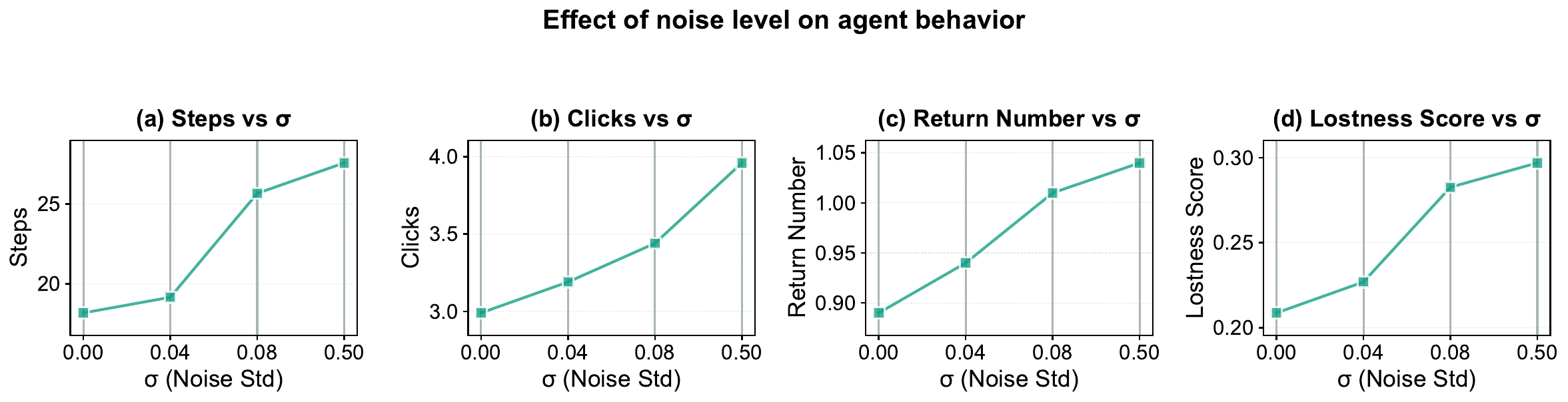}
    \caption{\rv{
    \emph{Effect of the noise level~$\sigma$ on agent behavior.} Panels show how increasing~$\sigma$ influences
(a) steps, (b) clicks, (c) return number, and (d) lostness score.
    }}
    \label{fig:noisestd}
\end{figure}

\begin{figure}[t]
    \centering
    \includegraphics[width=\textwidth]{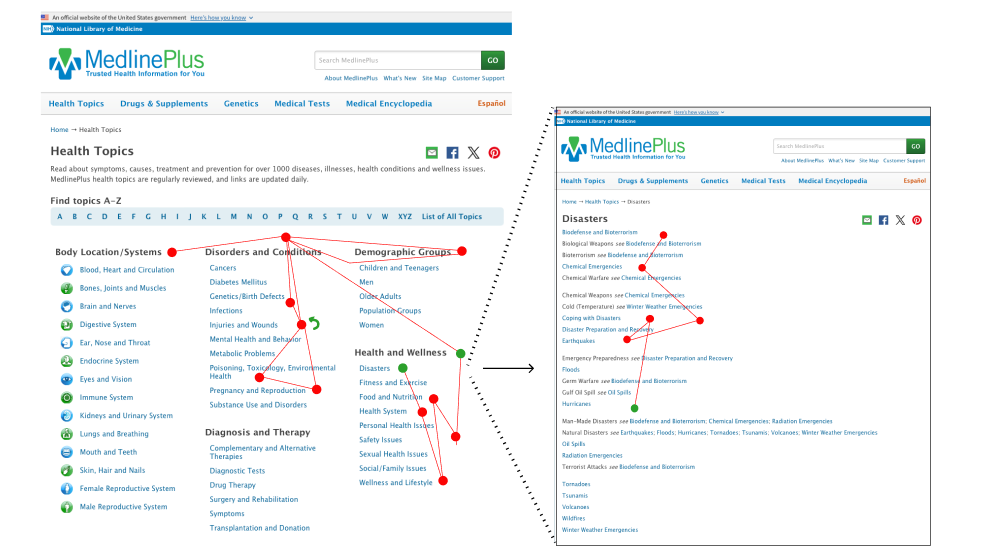}
    \vspace{6pt}
    \includegraphics[width=\textwidth]{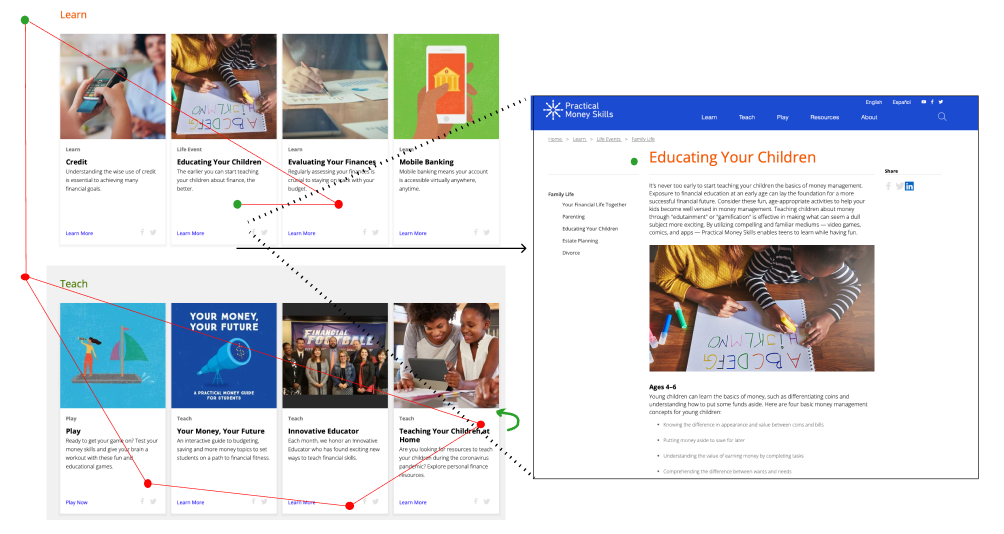}
    \caption{\rv{
    \emph{Agent trajectories on real webpages:}  (top) Case 1; (bottom) Case 2.  Dots indicate locations visited by the 
agent; curved arrows denote return actions taken after selecting or 
exploring an option. Green dots mark successful final selections.}
    }
    \label{fig:case}
\end{figure}

\rv{\section{Supplementary Tables and Figures}}
\rv{This section provides the following tables and figures.}
\begin{itemize}
    \item \rv{Table ~\ref{tab:variable} shows the summary of varibales used throughout the model formulation, equations, and experiments.}
    \item \rv{Figure~\ref{fig:sensi-detailed} shows key paparameters' detailed sensitivity on the three effects.}
    \item \rv{Figure~\ref{fig:memory-ablation} shows navigation behaviors with different memory decay threshold~$\theta$.}
    \item \rv{Figure~\ref{fig:noisestd} shows navigation behaviors with different noise level~$\sigma$.}
    \item \rv{Figure~\ref{fig:case} shows the examples that our agents on the real interfaces.}
    
\end{itemize}

\begin{table}[hb]
\centering
\caption{\rv{Summary of variables used throughout the model formulation, equations, and experiments.}}
\label{tab:variable}
\footnotesize
\begin{tabular}{lll}
\toprule
\textbf{Variable} & \textbf{Explanation} & \textbf{Reference} \\ 
\midrule

$M = \langle S, A, T, O, Z, R, \gamma \rangle$ & POMDP tuple (state, action, transition, observation, etc.) & Sec.~3.2 \\

$S$ & Full environment state & Sec.~3.2 \\
$A$ & Action set: \{\textsc{Visit}, \textsc{Select}, \textsc{Return}\} & Sec.~3.2 \\
$T$ & Transition function & Sec.~3.2 \\
$O$ & Observation space & Sec.~3.2 \\
$Z$ & Observation likelihood & Sec.~3.2 \\
$R$ & Reward function & Sec.~3.2 \\
$\gamma$ & Discount factor for future rewards & Eq.~(1), Sec.~5.5.3 \\

$s$ & Generic state in Bellman equation & Eq.~(1) \\
$a$ & Generic action in Bellman equation & Eq.~(1) \\
$r_t$ & Immediate reward at time $t$ & Eq.~(1) \\
$s_{t+1}$ & Next state & Eq.~(1) \\

$\pi$ & Policy & Eq.~(1)--(2) \\
$\pi^\ast$ & Optimal policy under utility–cost trade-off & Eq.~(2) \\

$U(s,a)$ & Task utility & Eq.~(2) \\
$C(s,a;\rho,t)$ & Cognitive cost function & Eq.~(2) \\
$\rho$ & Resource-limit parameters & Eq.~(2) \\
$t$ & Elapsed time & Eq.~(2) \\

\midrule
\multicolumn{3}{l}{\textbf{Information Scent Variables}} \\
\midrule

$\hat{\psi}_i$ & True information scent (semantic similarity) & Eq.~(3) \\
$\tilde{\psi}_i$ & Noisy observed scent & Eq.~(4) \\
$g$ & Goal embedding vector & Eq.~(3) \\
$l_i$ & Label embedding of option $i$ & Eq.~(3) \\
$\sigma$ & Std. dev. of Gaussian noise in scent & Eq.~(4), Table~1 \\

\midrule
\multicolumn{3}{l}{\textbf{Local Observation Panel}} \\
\midrule

$x^{\mathrm{loc}}_i$ & Local observation tuple $(\tilde{s}_i, v_i, c_i)$ & Sec.~3.2 \\
$v_i$ & Discretized visit count & Sec.~3.2 \\
$c_i$ & Discretized click count & Sec.~3.2 \\
$L_t$ & Set of options visible in current layer & Sec.~3.2 \\
$N_{\max}$ & Max list length used for padding (default 12) & Sec.~3.2 \\

\midrule
\multicolumn{3}{l}{\textbf{Global Memory Panel}} \\
\midrule

$M_i$ & Memory strength of option $i$ & Eq.~(5) \\
$q_i$ & Priority score: $\tilde{s}_i M_i$ & Sec.~3.2 \\
$\mathcal{G}(t)$ & Top-$K$ global memory items & Sec.~3.2 \\
$y^{\mathrm{glob}}_i$ & Global memory vector $(\tilde{s}_i, d^{\mathrm{norm}}_i)$ & Sec.~3.2 \\

$d^{\mathrm{norm}}_i$ & Normalized distance-to-goal & Sec.~3.2 \\
$\text{cost}(s_t \to i)$ & Shortest action-path cost to node $i$ & Sec.~3.2 \\
$d_{\max}$ & Max normalization distance & Sec.~3.2 \\

$\Delta k_i(t)$ & Steps since option $i$ was last viewed & Eq.~(5) \\
$V_i(t)$ & Cumulative visit count (long-term) & Eq.~(5) \\
$C_i(t)$ & Cumulative click count (long-term) & Eq.~(5) \\

$b$ & Baseline memory strength & Table~1 \\
$a_s$ & Weight of scent strength in memory & Table~1 \\
$a_v$ & Weight of visit frequency in memory & Table~1 \\
$a_c$ & Weight of clicks/selection in memory & Table~1 \\
$\lambda$ & Memory-decay rate ($H=5$ half-life) & Eq.~(5), Table~1 \\
$\theta$ & Retrieval threshold for forgetting & Table~1 \\

$K$ or $K_{\text{glob}}$ & Capacity of global-memory panel & Table~1 \\

\bottomrule
\end{tabular}
\end{table}

\end{document}